\documentclass{article}
\usepackage{amsmath}
\usepackage{amssymb}
\usepackage[all]{xy}
\usepackage{graphicx}
\newtheorem{defi}{Definition}[section]
\newtheorem{rem}{Remark}[section]
\newtheorem{thm}{Theorem}[section]

\newtheorem{lemma}[thm]{Lemma}
\newtheorem{cor}[thm]{Corollary}
\newtheorem{prop}[thm]{Proposition}
\newtheorem{conj}[thm]{Conjecture}
\newtheorem{example}[thm]{Example}

\newcounter{seq}
\newcounter{seqseq}
\newcommand\qed{\hfill\blacksquare\smallskip}

\newcommand\ee{\vect{e}}

\newcommand\PP{\ensuremath{\mathbb{P}}}
\newcommand\ZZ{\ensuremath{\mathbb{Z}}}
\newcommand\CC{\ensuremath{\mathbb{C}}}

\newcommand\RR{\ensuremath{\mathbb{R}}}

\newcommand\vect[1]{\mbox{\boldmath$#1$}}

\newcommand\ep{\varepsilon}

\newcommand\val{\mathrm{UD}}
\newcommand\Val{\mathrm{UD}}
\newcommand\zet[1]{\left\vert {#1} \right\vert}
\newcommand\kakko[1]{\left\vert\left\vert {#1} \right\vert\right\vert}
\newcommand\proof{\noindent{\textbf {Proof.}}\ \,}

\renewcommand\tilde{\widetilde}

\newcommand\ud{\mathop{-\!\triangleright}}

\title{Two dimensional periodic box-ball system and its fundamental cycle}

\author{Shinsuke Iwao\\
Graduate School of Science, Rikkyo University, \\
3-34-1 Nishi-Ikebukuro,
Toshima-ku, Tokyo, JAPAN}

\begin{document}

\maketitle

\begin{abstract} 
We study a $2$-dimensional Box-Ball system which is a ultradiscrete analog of the discrete KP equation. We construct an algorithm to calculate the fundamental cycle, which is an important conserved quantity of the $2$-dim.~Box-Ball system with periodic boundary condition, by using the tropical curve theory.
\end{abstract}

\section{Introduction}\label{sec1}

Ultradiscretization is a limiting procedure by which one can obtain piecewise linear equations from algebraic equations. This procedure allows us to make various piecewise linear dynamical systems, that are called \textit{ultradiscrete integrable systems}, from known discrete integrable systems. The first example of ultradiscrete integrable systems is the Takahashi-Satsuma Box-Ball system (BBS), discovered in 1990 \cite{Takahashi}, which is a dynamical system of balls in a one dimensional array of boxes. The BBS shows both a feature of cellular automata and that of solitons. %The ultradiscretization was introduced in 1996 as a certain limiting procedure  which connects the BBS and integrable nonlinear wave equations \cite{Tokihiro}.\\

After the discovery of the BBS, many researchers discovered various kinds of variants of the BBS, such as the BBS with carrier \cite{Takahashi2}, the BBS with colored balls \cite{Tokihiro2} and several kinds of $2$-dimensional BBS (2dBBS) \cite{Inoue, Inoue2, Moriwaki}. 

Around the same time, another relation between the BBS and the solvable quantum model was recognized \cite{Hatayama}. Both the ultradiscretization and the solvable quantum models are regarded as origins of ultradiscrete integrable systems.

In this paper, we study a $2$-dimensional BBS (2dBBS) which is a ultradiscretization of the \textit{discrete KP equation} (dKP)
\begin{equation}\label{eq1.1}
-\delta_m f_{n,m}^{t+1}f_{n+1,m+1}^t+(1+\delta_m)f_{n+1,m}^tf_{n,m+1}^{t+1}
-f_{n,m+1}^tf_{n+1,m}^{t+1}=0,
\end{equation}
with a certain periodic boundary condition. Introducing new variables
\[
I_{n,m}^t:=(1+\delta_m)\,\frac{f_{n,m}^tf_{n,m+1}^{t+1}}{f_{n,m+1}^tf_{n,m}^{t+1}},\quad 
V_{n,m}^t:=\delta_m\,\frac{f_{n,m}^tf_{n+1,m+1}^{t}}{f_{n+1,m}^tf_{n,m+1}^{t}},
\] 
we can rewrite dKP as $\left\{
\begin{array}{ll}
I_{n,m+1}^t+V_{n,m}^{t+1}=I_{n+1,m}^t+V_{n,m+1}^t\\  
I_{n,m}^tV_{n,m}^{t+1}=I_{n+1,m}^tV_{n,m}^t          
\end{array}
\right.$. Further, we can transform this system as follows:
\begin{lemma}\label{lem:1.1}
The following system implies dKP:
\begin{gather}
\textstyle
I_{n,m}^t=V_{n,m}^t\!\left\{
1+
{
\left[\frac{V_{n,m}^t}{I_{n+1,m-1}^t}\!+\!\frac{V_{n,m}^tV_{n,m-1}^t}{I_{n+1,m-1}^tI_{n+1,m-2}^t}\!+\!
\frac{V_{n,m}^tV_{n,m-1}^tV_{n,m-2}^t}{I_{n+1,m-1}^tI_{n+1,m-2}^tI_{n+1,m-3}^t}
+\!\cdots
\right]^{-1}}
\right\}\label{eq:2}\\
I_{n,m}^tV_{n,m}^{t+1}=I_{n+1,m}^tV_{n,m}^t.\label{eq:3}
\end{gather}
\end{lemma}

\proof Let $D_m:=\frac{V_{n,m}^t}{I_{n+1,m-1}^t}+\frac{V_{n,m}^tV_{n,m-1}^t}{I_{n+1,m-1}^tI_{n+1,m-2}^t}+\frac{V_{n,m}^tV_{n,m-1}^tV_{n,m-2}^t}{I_{n+1,m-1}^tI_{n+1,m-2}^tI_{n+1,m-3}}+\!\cdots$. Then $D_m$ satisfies $D_m=\frac{V_{n,m}^t}{I_{n+1,m-1}^t}(1+D_{m-1})$. This equation and (\ref{eq:2}--\ref{eq:3}) imply $I_{n,m+1}^t=V_{n,m+1}^t(1+D_{m+1}^{-1})=V_{n,m+1}^t+I_{n+1,m}^t-V_{n,m}^{t+1}$. $\qed$

Under the periodic boundary condition $I_{n,m}^t\equiv I_{n+N,m}^t\equiv I_{n,m+M}^t$, $V_{n,m}^t\equiv V_{n+N,m}^t\equiv V_{n,m+M}^t$, the quantities $\textstyle\alpha=\prod_{m=1}^M{I_{n,m}^t}=\prod_{m=1}^M{(1+\delta_m)}, \quad\beta=\prod_{m=1}^M{V_{n,m}^t}=\prod_{m=1}^M{\delta_m}$ should be invariant under the shifts $n\mapsto n+1$, $t\mapsto t+1$. Assume $\alpha>\beta>0$. Then, (\ref{eq:2}) is equivalent to 
\begin{equation}\label{eq:5}
I_{n,m}^t=V_{n,m}^t\!\left(\!
1\!+\!
\frac{1-\beta/\alpha}
{\frac{V_{n,m}^t}{I_{n+1,m-1}^t}+\frac{V_{n,m}^tV_{n,m-1}^t}{I_{n+1,m-1}^tI_{n+1,m-2}^t}
+\!\cdots\!+\frac{V_{n,m}^tV_{n,m-1}^t\cdots V_{n,m+1}^t}
{I_{n+1,m-1}^tI_{n+1,m-2}^t\cdots I_{n+1,m}^t}}
\!\right).
\end{equation}
If all $I_{n,m}^t$, $V_{n,m}^t$ are positive, we introduce a new positive parameter $\ep>0$ and the variable transformation: $I_{n,m}^t=e^{-\frac{Q_{n,m}^t}{\ep}}$ $V_{n,m}^t=e^{-\frac{W_{n,m}^t}{\ep}}$, $\alpha=k_1 e^{-\frac{A}{\ep}}$, $\beta=k_2 e^{-\frac{B}{\ep}}$. The 2dBBS is the following piecewise linear system which is obtained by taking the limit $\ep\to 0^+$ of the equations (\ref{eq:3}, \ref{eq:5}):
\begin{gather}
\left\{
\begin{array}{l}
\textstyle Q_{n,m}^{t}=W_{n,m}^t+\min\left[0,\ X_{n,m}^t\right],\quad
\textstyle W_{n,m}^{t+1}=Q_{n+1,m}^t+W_{n,m}^t-Q_{n,m}^{t}\\
X_{n,m}^t:=\textstyle\max_{k=0}^{M-1}
{[\sum_{l=0}^k{(Q_{n+1,m-l-1}^t-W_{n,m-l}^t)}]}
\end{array}
\right..
\label{eq:6}
\end{gather}
We note $A=\sum_{m=1}^M{Q_{n,m}^t}$, $B=\sum_{m=1}^MW_{n,m}^t$.

\subsection{Examples of 2dBBS}

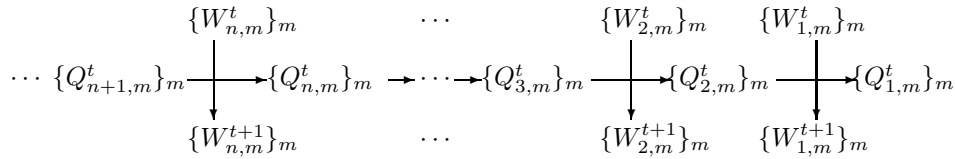
\begin{figure}[htbp]
\begin{center}
\begin{picture}(340,50)(-58,0)
%---------------------------------------
\put(-42,23){$\{{Q}_{n+1,m}^t\}_m$}
\put(39,23){$\{{Q}_{n,m}^t\}_m$}
\put(120,23){$\{{Q}_{3,m}^t\}_m$}
\put(190,23){$\{{Q}_{2,m}^t\}_m$}
\put(260,23){$\{{Q}_{1,m}^t\}_m$}
%----------------------------------------------
\put(9,45){$\{{W}_{n,m}^t\}_{m}$}
\put(9,0){$\{{W}_{n,m}^{t+1}\}_m$}
\put(165,45){$\{{W}_{2,m}^t\}_{m}$}
\put(165,0){$\{{W}_{2,m}^{t+1}\}_m$}
\put(225,45){$\{{W}_{1,m}^t\}_{m}$}
\put(225,0){$\{{W}_{1,m}^{t+1}\}_m$}
%------------------------------------------
\put(9,25){\vector(1,0){30}}
\put(85,25){\vector(1,0){10}} 
\put(-58,23){$\cdots$}
\put(97,23){$\cdots$}
\put(97,45){$\cdots$}
\put(97,0){$\cdots$}
\put(110,25){\vector(1,0){10}}
\put(162,25){\vector(1,0){30}}
\put(232,25){\vector(1,0){30}}
%-------------------------------------------------
\put(19,40){\vector(0,-1){30}}
\put(177,40){\vector(0,-1){30}}
\put(247,40){\vector(0,-1){30}}
\end{picture}
\end{center}
\caption{The time evolution rule of 2dBBS.}
\label{fig:1}
\end{figure}

We can regard 2dBBS as a dynamical system $\{W_{n,m}^t\}_{n,m}\mapsto\{W_{n,m}^{t+1}\}_{n,m}$. (See figure \ref{fig:1}).
Although the existence of $\{Q_{n,m}^t\}_{n,m}$ satisfying the periodic boundary condition $Q_{n+M,m}^t=Q_{n,m}^t$ is non-trivial, we can prove the following statement:
\begin{prop}\label{prop1.2}
For any $\{W_{n,m}^t\}_{n,m}\in \RR^{N\times M}$ and real number $A$, there uniquely exists $\{Q_{n,m}^t\}_{n,m}$ satisfying the periodic boundary condition and $A=\sum_{m=1}^M{Q_{n,m}^t}$ $\forall n$.
\end{prop}
\proof We give the proof in the next section (Remark \ref{rem2.3}). $\qed$

We express the state of the 2dBBS by laying out the following data on two dimensional space: (i) $N\times M$ numbers $\{W_{n,m}^t\}_{n,m}$,(ii) $M$ numbers $\{Q_{1,m}^t\}_m$ as figure \ref{fig:2}. This expression is similar to the graphical interpretation of the 2dBBS which is introduced in \cite{Inoue,Inoue2}.

\begin{figure}[htbp]
\begin{center}
\begin{picture}(200,80)(-40,-10)
%----------------------------------------------
\put(-40,55){\framebox(25,20){$Q_{1,M}^t$}}
\put(-40,40){\framebox(25,15){$\vdots$}}
\put(-40,20){\framebox(25,20){$Q_{1,2}^t$}}
\put(-40,0){\framebox(25,20){$Q_{1,1}^t$}}
%----------------------------------------------
\put(0,55){\framebox(25,20){$W_{N,M}^t$}}
\put(0,40){\framebox(25,15){$\vdots$}}
\put(0,20){\framebox(25,20){$W_{N,2}^t$}}
\put(0,0){\framebox(25,20){$W_{N,1}^t$}}
%-----------------------------------
\put(25,55){\framebox(20,20){$\cdots$}}
\put(25,40){\framebox(20,15){}}
\put(25,20){\framebox(20,20){$\cdots$}}
\put(25,0){\framebox(20,20){$\cdots$}}
%%-----------------------------------
\put(45,55){\framebox(25,20){$W_{3,M}^t$}}
\put(45,40){\framebox(25,15){$\vdots$}}
\put(45,20){\framebox(25,20){$W_{3,2}^t$}}
\put(45,0){\framebox(25,20){$W_{3,1}^t$}}
%%-----------------------------------
\put(70,55){\framebox(25,20){$W_{2,M}^t$}}
\put(70,40){\framebox(25,15){$\vdots$}}
\put(70,20){\framebox(25,20){$W_{2,2}^t$}}
\put(70,0){\framebox(25,20){$W_{2,1}^t$}}
%%-----------------------------------
\put(95,55){\framebox(25,20){$W_{1,M}^t$}}
\put(95,40){\framebox(25,15){$\vdots$}}
\put(95,20){\framebox(25,20){$W_{1,2}^t$}}
\put(95,0){\framebox(25,20){$W_{1,1}^t$}}
%-----------------------------------
\put(-10,-10){$n=N$}
\put(27.5,-10){$\cdots$}
\put(55,-10){$3$}
\put(80,-10){$2$}
\put(105,-10){$1$}
%------------------------------------
\put(125,61){$M$}
\put(128,41){$\vdots$}
\put(128,26){$2$}
\put(128,6){$1=m$}
\end{picture}
\end{center}
\caption{The state of 2dBBS at time $t$.}
\label{fig:2}
\end{figure}

The following example shows the collision of two solitons ($A=1,B=3$).

\begin{verbatim}
     .|...1.2...      .|....111..     .|.....2.1.     .|.....11.1
     .|..1.2....      .|...12....     .|....3....     .|....21...
     .|.1.2.....      .|..12.....     .|...3.....     .|...21....
     1|322.11333      1|332..2233     1|333..1323     1|3331.1232
          t=0             t=1             t=2             t=3
\end{verbatim}

This paper is arranged as follows: In \S \ref{sec:2}, we review the connection between the dKP and the algebraic geometry. We introduce the \textit{spectral curve} of the dKP, that is an important invariant of the equation. In \S \ref{sec:3}, we review several basic results of the tropical geometry according to \cite{Iwao,Tropical1,Mikhalkin,Tropical2}. In \S \ref{sec4}, we obtain a tropical analog of the spectral curve of 2dBBS and a tropical-geometric expression of the invariants. Moreover, we have an algorithm to calculate the fundamental cycle of 2dBBS. 

{\bf Notations:} For a positive integer $M$
and a ring $R$ with $1$, $\mathrm{Mat}(M,R)$ denotes the $R$-algebra of matrices over $R$ of size $M$. $E$ denotes the identity matrix and  $E_{i,j}$ denotes the matrix of which $(i,j)$-element is $1$, and other elements are $0$. We denote by $\mathrm{diag}(a_1,\dots,a_M)$ the diagonal matrix of which $(i,i)$-element equals to $a_i$. Define $S:=(\delta_{i+1,j})_{i,j}+y\cdot E_{M,1}\in\mathrm{Mat}(M,\CC[y])$.

Let $\mathcal{O}$ be a $\CC[y^{\pm 1}]$-algebra and $\mathcal{O}^\infty[M]$ be the $\mathcal{O}$-module defined by 
\[
\mathcal{O}^\infty[M]:=\{(\cdots,a_{-1},a_0,a_1,\dots)\,\vert\,a_i\in \mathcal{O}, a_{i+M}=y\cdot a_i\}, 
\]
which is isomorphic to $\mathcal{O}^M= \mathcal{O}\times\cdots\times \mathcal{O}$ ($M$ times) as $\mathcal{O}$-modules through the isomorphism $\mathcal{O}^\infty[M]\to \mathcal{O}^M$; $(a_i)_{i\in\ZZ}\mapsto (a_i)_{i=1}^M$. An element of $\mathcal{O}^\infty[M]$ is called a $M$-\textit{periodic vector}. 

The set of $\mathcal{O}$-automorphisms of $\mathcal{O}^M$ is naturally identified as $\mathrm{Mat}(M,\mathcal{O})$. Similarly, the set of $\mathcal{O}$-automorphisms of $\mathcal{O}^\infty[M]$ is identified as the $\mathcal{O}$-submodule $P$ of $\mathrm{Mat}(\infty,\CC)$ which is the image of the morphism
\[\mathrm{Mat}(M,\CC[y^{\pm1}])\to \mathrm{Mat}(\infty,\CC);\quad(z_{i,j})_{i,j=1}^M\mapsto (\tilde{z}_{i,j})_{i,j\in\ZZ},\] 
where 
$z_{i,j}=\sum_{n\in\ZZ}{\tilde{z}_{i,j+nM}\cdot y^{n}} \quad (1\leq i,j\leq M)$, $\tilde{z}_{i+M,j}=\tilde{z}_{i,j}$.
An element of $P$ is called a \textit{$M$-periodic matrix}. For a matrix $Z=(z_{i,j})_{i,j=1}^M\in \mathrm{Mat}(M,\mathcal{O})$, we express the associated $M$-periodic matrix as $\tilde{Z}=(\tilde{z}_{i,j})_{i,j\in\ZZ}$. Similarly, for $a=(a_i)_{i=1}^M\in \mathcal{O}^M$, we denote the associated $M$-periodic vector by $\tilde{a}=(\tilde{a}_i)_{i\in\ZZ}$. It follows that $(\tilde{Z\cdot a})=\tilde{Z}\cdot \tilde{a}$.

\section{Periodic discrete KP equation}\label{sec:2}

\subsection{spectral curve}\label{sec3.1}

Periodic discrete KP equation (Lemma \ref{lem:1.1})
is rewritten as the following matrix form:

\begin{equation}\label{eq:10}
L_n^{t+1}(y)R_{n}^t(y)=R_{n+1}^t(y)L_n^t(y),
\end{equation}
where $L_n^t(y)=\mathrm{diag}(V_{n,1}^t,\dots,V_{n,M}^t)+S$, $R_n^t(y)=\mathrm{diag}(I_{n,1}^t,\dots,I_{n,M}^t)+S$. By using the new matrix
\begin{equation}\label{eq:11}
X_n^t(y):=L_{n+N-1}^t\cdots L_{n+1}^tL_n^t,\qquad (L_n^t=L_n^t(y)),
\end{equation}
we can transform the dKP to the following matrix form:
\begin{equation}\label{eq:12}
X_n^{t+1}R_n^t=R_n^{t}X_n^t,\quad
\mbox{or equivalently,}\quad X_{n+1}^tL_n^t=L_n^tX_{n}^t.
\end{equation}

Form (\ref{eq:12}), the characteristic polynomial $\Phi(x,y):=\det{(X_n^t(y)-xE)}$ should be invariant under $t\mapsto t+1$, $n\mapsto n+1$. Let $\tilde{C}\subset \PP^1\times\PP^1$ be the algebraic curve defined by $\Phi$ and $C_0\subset \CC^2$ be the affine part of $\tilde{C}$. Put $\eta=x^{-1}$, $\zeta=y^{-1}$, $d:=\mathrm{g.c.d.}(N,M)$, $N=dN_1$ and $M=dM_1$. 
By direct calculations, we have the following expression \cite{Tokihiro3}:
\begin{equation}\label{eq:13}
\eta^M\zeta^N\cdot\Phi=(\eta^{M_1}-\zeta^{N_1})^d+\sum_{NM<Ni+Mj\leq 2NM,\ %\atop
i,j\geq 0}{c_{i,j}\,\eta^i\zeta^j},\qquad
c_{i,j}\neq 0,
\end{equation}
which implies 
that the set $\tilde{C}\setminus C_0$ 
consists of only one point 
${p}_0:(\eta,\zeta)=(0,0)$.

By (\ref{eq:13}), we can introduce a coordinate variable $k$ such that $\eta=k^{N}\cdot(1+c_1k+c_2k^2+\cdots)$ and $\zeta=k^{M}$, where $c_1\neq 0$. Denote by $C$ the analytic curve obtained from $\tilde{C}$ of which the local coordinate at $p_0$ is $k$. We call $C$ a \textit{spectral curve} of the dKP. 

Let us define the following two sets: 
\begin{gather*}
 \textstyle\mathcal{V}(\beta):=\{(V_{n,m})_{n,m}\in\CC^{N\times M}\,\vert\,\beta=\prod_m{V_{n,m}\ (\forall n)}\},\\
\mathcal{W}(\beta):=\left\{X \,\left\vert\, X=L_N\cdots L_2L_1,\ L_n=\mathrm{diag}(V_{n,1},\dots,V_{n,M})+S,\ (V_{n,m})\in\mathcal{V}(\beta)\right\}.\right.
\end{gather*}
Let $j:\mathcal{V}(\beta)\to \mathcal{W}(\beta)$ be the natural surjection which sends an element $(V_{n,m})_{n,m}\in \mathcal{V}(\beta)$ to the matrix $X=L_N\cdots L_2L_1$, $L_n=\mathrm{diag}(V_{n,1},\dots,V_{n,M})+S$. One can prove that $j$ is bijective (Lemma 3.1. \cite{kajiwara}). We will often identify $\mathcal{V}(\beta)$ with $\mathcal{W}(\beta)$.

Through this procedure, we have the following map:
\begin{equation}\label{eq:14}
\mathrm{SC}:\ \mathcal{V}(\beta)\cong\mathcal{W}(\beta)\to \{\mbox{spectral curves of dKP}\}.
\end{equation}
The \textit{isolevel set associated with a spectral curve $C$} is the inverse image $\mathcal{T}_C:=\mathrm{SC}^{-1}(C)$.

For two matrices $X,Y\in \mathcal{T}_C$, we say $X\sim Y$ if there exists an invertible diagonal matrix $D$ such that $X=DYD^{-1}$. This defines an equivalence relation over $\mathcal{T}_C$. We write the quotient set induced by this relation as $\mathcal{R}_C:=\mathcal{T}_C/\sim$.

\subsection{time evolution of dKP}\label{sec2.2}

Unfortunately, the equation $X_{n}^{t+1}R_n^t=R_n^tX_n^t$ (\ref{eq:12}) does not define the time evolution $X_{n}^t\mapsto X_n^{t+1}$ uniquely. In fact, the following proposition can be proved.
\begin{prop}\label{prop:2.1}
For given $X^t_n\in\mathcal{W}(\beta)$, there exist $M$ pairs $(R^{(1)},X^{(1)}),\cdots$, $(R^{(M)},X^{(M)})$ such that $(i)$ $R^{(i)}=\mathrm{diag}(I^{(i)}_{1},\dots,I^{(i)}_M)+S$, $(ii)$ $\alpha=\prod_{m=1}^M{I_m^{(i)}}$, $(iii)$ $X^{(i)}\in\mathcal{W}(\beta)$, $(iv)$ $X^{(i)}R^{(i)}=R^{(i)}X_n^t$.
\end{prop}
However, we can uniquely specify a certain time evolution with an additional condition.
\begin{prop}\label{prop:2.2}
Let $X_n^t=L_{n+N-1}\cdots L_{n+1}L_n\in\mathcal{W}(\beta)$, $L_n^t=\mathrm{diag}(V_{n,1}^t,\dots,V_{n,M}^t)+S$ such that $V_{n,m}^t>0$. Then, there uniquely exist two matrices $X_n^{t+1}$, $R_n^t$ such that $(i)$ $R_n^t=\mathrm{diag}(I^t_{1},\dots,I^t_M)+S$, $(ii)$ $\alpha=\prod_{m=1}^M{I_m^t}$, $(iii)$ $X_n^{t+1}\in\mathcal{W}(\beta)$, $(iv)$ $X_n^{t+1}R_n^t=R_n^tX_n^t$, $(v)$ $I_{n,m}^t>0$.
\end{prop}

By this proposition, we can say that a certain type of the time evolution of $X_n^t$ is determined by (\ref{eq:12}).
%
%\begin{defi}
%For positive real dependent variables $\{V_{n,m}^t\}_{n,m}$,
%we call
%the time evolution $X_n^t\mapsto X_n^{t+1}=R_n^tX_n^t(R_n^t)^{-1}$
%which is obtained in proposition \ref{prop:2.2}
%the principal time evolution.
%Through analytic continuation, the principal time evolution is extended 
%to any complex dependent variables $\{V_{n,m}^t\}_{n,m}$.
%\end{defi}
We will give the proofs of propositions 
\ref{prop:2.1}, \ref{prop:2.2}
in the appendix.

\begin{rem}\label{rem2.3}
Proposition $\ref{prop:2.1}$ implies
proposition $\ref{prop1.2}$.
\end{rem}

\subsection{linearisation}\label{sec:2.3}

The method to solve the initial value problem of discrete integrable systems with a spectral curve have been established by many authors. We review some fundamental results concerning the \textit{linearisation}. %of the integrable systems}.

We consider the three automorphisms $T_t,T_n,T_m$ of $\mathcal{T}_C$ defined by
\begin{gather}
 T_t(X_n^t):=X_n^{t+1}=R_n^tX_n^t(R_n^t)^{-1},\quad
T_n(X_n^t):=X_{n+1}^t=L_n^tX_n^t(L_n^t)^{-1},\\
T_m(X_n^t):=X_n^{t}\vert_{m\mapsto m+1}=SX_n^tS^{-1},
\end{gather}
where $T_t$ is the unique time evolution defined in \S\ref{sec2.2}. Let $X_{n,m}^t:=S^{m-1}X_n^tS^{-m+1}$. Then it follows that $X_{n,m+1}^t=T_m(X_{n,m}^t)$.

For an algebraic curve $C$, we write $\mathrm{Div}\,C:=\bigoplus_{p\in C}{\ZZ\cdot p}$. The Picard group $\mathrm{Pic}\,C$ is the quotient group of $\mathrm{Div}\,C$ induced by the linear equivalence relation. Denote by $\mathrm{Pic}^n\, C$ the subset of $\mathrm{Pic}\, C$ consisted of the divisors of degree $n$. For a rational function $f$ over $C$, $(f)_0$ (\textit{resp.~}$(f)_\infty$) denotes the divisor of zeros (\textit{resp.~}poles) of $f$. The following theorem \ref{thm:2.3} is a fundamental result concerning the algebro-geometric aspects of spectral curves proved in \cite{Moerbeke}. 

\begin{thm}\label{thm:2.3}
There exists a mapping 
$\tilde{\varphi}:\mathcal{R}_C\to \mathrm{Div}^g\, C$
such that:\\
$(i)$ $\tilde{\varphi}(X)$ is a general divisor of degree $g$ for any $X\in\mathcal{R}_C$, \\
$(ii)$ the induced mapping $\varphi:\mathcal{R}_C\to \mathrm{Pic}^g\, C$ is injective,\\
$(iii)$ we have $\tilde{\varphi}(X)=(f_1/f_M)_0-(M-1)p_0$, where $f_i=f_i(x,y)$ is the $i$-th component of a eigenvector of $X=X(y)$ belonging to an eigenvalue $x$.
$\hfill\blacksquare$
\end{thm}
%Note that $f_i$ should be a rational function over $C$. (Cramer's rule)

Let $X_n^t\in\mathcal{T}_C$. We denote by $\vect{v}_n^t=(f_{n,1}^t,\dots,f_{n,M}^t)^T$ a eigenvector of $X_n^t$ belonging to $x$ \textit{i.e.} 
\begin{equation}\label{eq:18}
X_n^t\vect{v}_n^t=x\, \vect{v}_n^t.
\end{equation}
By (\ref{eq:12}),
we have the following evolution equation
\begin{equation}\label{eq:19}
\vect{v}_{n}^{t+1}=R_n^t\vect{v}_n^t,\qquad
\vect{v}_{n+1}^t=L_n^t{\vect{v}}_n^t.
\end{equation}

Equations (\ref{eq:18}, \ref{eq:19}) can be regarded as linear equations over the rational function field $\mathcal{K}$ over $C$. Naturally, $\mathcal{K}$ has a structure of $\CC[y^{\pm 1}]$-algebra. Using $M$-periodic matrices and vectors, we rewrite these equations as 
\begin{equation}\label{eq:20}
\tilde{X}_n^t\tilde{\vect{v}}_n^t=x\, \tilde{\vect{v}}_n^t,\qquad
\tilde{\vect{v}}_{n}^{t+1}=\tilde{R}_n^t\tilde{\vect{v}}_n^t,\qquad
\tilde{\vect{v}}_{n+1}^t=\tilde{L}_n^t\tilde{{\vect{v}}}_n^t.
\end{equation}

By definition of $X_n^t$, the $M$-periodic matrix $\tilde{X}_n^t$ is expressed as $\tilde{X}_n^t=\tilde{S}^{N}+\tilde{Z}$, where $\tilde{Z}=(\tilde{z}_{i,j})_{i,j}$ is a $M$-periodic matrix such that $\tilde{z}_{i,j}\neq 0 \Rightarrow 0\leq j-i< N$. Then, the first equation of (\ref{eq:20}) is rewritten as $\tilde{S}^{N}\tilde{\vect{v}}_n^t=(x\tilde{E}-\tilde{Z})\tilde{\vect{v}}_n^t$. Equivalently, we have 
\begin{equation}\label{eq:21}
\textstyle
\tilde{f}_{n,m+N}^t=x\tilde{f}_{n,m}^t-\sum_{k=0}^{N-1}{\tilde{z}_{m,m+k}\,
\tilde{f}_{n,m+k}^t}.
\end{equation}

Let $\vect{w}_n^t=(\tilde{f}_{n,1}^t,\dots. \tilde{f}_{n,N}^t)^T$, $M_n^t=\mathrm{diag}(I_{n,1}^t,\dots,I_{n,N}^t)+U_1$ and $H_n^t=\mathrm{diag}(V_{n,1}^t,\dots,V_{n,N}^t)+U_1$, where $U_m=(\delta_{i+1,j})_{i,j}+(x-\tilde{z}_{m,m})E_{N,1}-\tilde{z}_{m,m+1}\cdot E_{N,2}-\cdots-\tilde{z}_{m,m+N-1}\cdot E_{N,N}$. Then, from (\ref{eq:21}), we can rewrite (\ref{eq:19}) as 
\begin{equation}\label{11301226}
\vect{w}_{n}^{t+1}=M_n^t\vect{w}_n^t,\qquad \vect{w}_{n+1}^t=H_n^t{\vect{w}}_n^t.
\end{equation}

\begin{thm}[\cite{Iwao2}]\label{thm:2.4}
$(i)$  There exist divisors $\mathcal{T}$, $\mathcal{N}$, $\mathcal{M}_1$, $\mathcal{M}_2$, $\dots,\mathcal{M}_M\in\mathrm{Pic}(C)$ of degree $0$ such that $\varphi(X_{n,m}^{t+1})=\varphi(X_{n,m}^t)+\mathcal{T}$, $\varphi(X_{n+1,m}^{t})=\varphi(X_{n,m}^t)+\mathcal{N}$, $\varphi(X_{n,m+1}^{t})=\varphi(X_{n,m}^t)+\mathcal{M}_m$.\\
$(ii)$ 
The divisors $\mathcal{T}$, $\mathcal{N}$, $\mathcal{M}_m$
are expressed as follows:
\begin{equation}
\mathcal{T}=p_\mathcal{T}^+-p_\mathcal{T}^-,\quad
\mathcal{N}=p_\mathcal{N}^+-p_\mathcal{N}^-,\quad
\mathcal{M}_m=p_\mathcal{M}^+(m)-p_\mathcal{M}^-(m),
\end{equation}
where 
$p_\mathcal{T}^\pm$ %:(x_\mathcal{T}^\pm,y_\mathcal{T}^\pm)$, 
$p_\mathcal{N}^\pm$ %:(x_\mathcal{N}^\pm,y_\mathcal{N}^\pm)$, 
$p_\mathcal{M}^\pm(m)$ %:(x_\mathcal{M}^\pm(m),y_\mathcal{M}^\pm(m))$
are certain points on $C$.\\
$(iii)$
The $x,y$ coordinates of $p_\mathcal{T}^\pm$, $p_\mathcal{N}^\pm$, $p_\mathcal{M}^\pm(m)$ are expressed as
\begin{gather*}
x_\mathcal{T}^+=(\det{M_n^t})_\infty, \quad
x_\mathcal{T}^-=(\det{M_n^t})_0,\quad
y_\mathcal{T}^+=(\det{R_n^t})_\infty,\quad 
y_\mathcal{T}^-=(\det{R_n^t})_0,\\
%--------------------------------------------------
x_\mathcal{N}^+=(\det{H_n^t})_\infty, \quad
x_\mathcal{N}^-=(\det{H_n^t})_0,\quad
y_\mathcal{N}^+=(\det{L_n^t})_\infty,\quad 
y_\mathcal{N}^-=(\det{L_n^t})_0,\\
%------------------------------------------------------
x_\mathcal{M}^+(m)=(\det{U_m})_\infty, \quad
x_\mathcal{M}^-(m)=(\det{U_m})_0,\\
y_\mathcal{M}^+(m)=(\det{S})_\infty,\quad 
y_\mathcal{M}^-(m)=(\det{S})_0. \hspace{10pt}\mbox{$\qed$}
\end{gather*}
\end{thm}

Roughly speaking, theorem \ref{thm:2.4} states that the determinants of matrices $L_n^t$, $R_n^t$, $H_n^t$, $M_n^t,\cdots$ characterize the action of $T_n,T_m,T_t$. %On the other hand, by using the techniques in linear algebra, we can obtain the following results:

\begin{lemma}\label{lem:2.5}
$\det{R_n^t}=\alpha-(-1)^My$, $\det{L_n^t}=\beta-(-1)^My$, $\det{H_n^t}=(-1)^{N+1}x$, $\det{S}=(-1)^{M+1}y$, $\det{U_m}=(-1)^{N+1}(x-V_{N,m}^{t}\cdots V_{2,m}^{t}V_{1,m}^{t})$, $\det{M_n^t}=(-1)^N(x-\kappa)$, where $\kappa$ is a minimum eigenvalue of $X_n^t\vert_{y=(-1)^M\alpha}$ in absolute value.
\end{lemma}
\proof We prove these lemmas in the appendix. $\qed$

From theorem \ref{thm:2.4} and lemma \ref{lem:2.5}, the representations of the actions $T_t$, $T_n$, $T_m$ on $\mathrm{Pic}^g\,C$ are completely determined:
\begin{prop}\label{prop:2.6}
The divisors $\mathcal{T}$, $\mathcal{N}$, $\mathcal{M}_m$
in theorem \ref{thm:2.4} are expressed as 
$\mathcal{T}=p_0-p_1$, $\mathcal{N}=p_0-p_2$, $\mathcal{M}_m=p_0-p_3^{(m)}$,
where
$p_1:(x,y)=(\kappa,\alpha)$,
$p_2:(x,y)=(0,\beta)$, 
$p_3^{(m)}:(x,y)=(\gamma_m,0)$,
$\gamma_m=V_{N,m}^{t}\cdots V_{2,m}^{t}V_{1,m}^{t}$.
\end{prop}

\subsection{$\mathcal{T}_C$ and $\mathcal{R}_C$}\label{sec:2.4}

Theorem \ref{thm:2.4} states that the evolutions $t\mapsto t+1$, $n\mapsto n+1$, $m\mapsto m+1$ over $\mathcal{R}_C$ can be linearized on $\mathrm{Pic}^g(C)$ through the injection $\varphi:\mathcal{R}_C\to\mathrm{Pic}^g(C)$. To lift this result to the isolevel set $\mathcal{T}_C$, we will study the structures of the two sets $\mathcal{R}_C$ and $\mathcal{T}_C$. Let $d:=\mathrm{g.c.d.}(N,M)$.

Our aim in this section is to construct a bijection
\begin{equation}
\mathcal{T}_C\to\mathcal{R}_C\times (\CC^\times)^{d-1} 
\end{equation}
which has some nice behavior. Denote by $[X]$ the image of $X\in \mathcal{T}_C$ through the natural projection $\mathcal{T}_C\to \mathcal{R}_C=\mathcal{T}_C/\sim$. If $[X]=[Y]$, there exits a diagonal invertible matrix $D$ such that $X=DYD^{-1}$.
\begin{lemma}\label{lem:2.7}
Let $D$ be a diagonal and invertible matrix. Then, $X,DXD^{-1}\in \mathcal{T}_C\iff S^dDS^{-d}=D$.
\end{lemma}
\proof $\Rightarrow)$\
Because any element $X\in\mathcal{T}_C$ is written as $X=S^N+\sum_{i=0}^{N-1}{X_iS^i}$, ($X_i$ are diagonal) uniquely, we have $S^N DS^{-N}=D$. On the other hand, by definition of the matrix $S$, it follows that $S^M D S^{-M}=D$. Due to these equations, we have $S^dDS^{-d}=D$. $\Leftarrow)$\ Any element $X\in\mathcal{T}_C$ is written as $X=(L_N+S)\cdots(L_2+S)(L_1+S)$. Putting $L_n':=(S^{-n}DS^{n})L_n(S^{n-1}D^{-1}S^{-n+1})$ and $X':=(L_N'+S)\cdots(L_1'+S)$, we have $X'=DXD^{-1}\in\mathcal{T}_C$.
$\qed$

Let $G(\cong (\CC^\times)^d)$ be the multiplicative group of diagonal matrices defined by 
\[
G:=\{D\,\vert\,D:\mbox{invertible, diagonal,}\ S^dDS^{-d}=D\},
\] 
and $H(\cong \CC^\times)\subset G$ be the normal subgroup $H:=\{cE\,\vert\,c\in\CC^\times\}$. By lemma \ref{lem:2.7}, the quotient group $G/H$ acts on $\mathcal{T}_C$ by $G/H\ni\ D\mapsto \{X\mapsto DXD^{-1}\}\ \in \mathrm{End}(\mathcal{T}_C)$. Because the group action is free, the orbit $(G/H)\cdot X$ is isomorphic to $G/H\cong (\CC^\times)^{d-1}$ for any $X\in \mathcal{T}_C$.

Therefore, any map $i:\mathcal{T}_C\to G/H$ satisfying $i(DXD^{-1})=D\cdot i(X)$ induces a bijection $\mathcal{T}_C\to \mathcal{R}_C\times (G/H)$. To define a good bijection, we focus on a eigenvector of $X$ at the point $p_0\in C$. Let $k$ be the local coordinate around $p_0\in C$. 

\begin{lemma}\label{lem:2.8}
Let $X$ be an element of $\mathcal{T}_C$, $\mathcal{K}$ be the rational function field of $C$. Assume $\vect{v}\in\mathcal{K}^M$ satisfies $X\vect{v}=x\vect{v}$. Up to multiplication by a constant, the $i$-th component $f_i$ of $\vect{v}$ has the following expression near $p_0$:
\[
f_i=a_ik^{M-i}+o(k^{M-i}),\qquad a_i\neq 0.
\]
Moreover, we have $a_{i+d}=a_i$ for all $i$.
\end{lemma}

\proof
See the appendix.$\qed$

Define the map $i:\mathcal{T}_C\to G/H(\cong(\CC^\times)^d/\CC^\times)$ by $X\mapsto [a_1:a_2:\dots:a_d]^T$, where $a_i$ is the non-zero number in lemma \ref{lem:2.8}. By definition of $i$, we can check $i(DXD^{-1})=D\cdot i(X)$ immediately. As stated above, this map induces a bijection $\mathcal{T}_C\to \mathcal{R}_C\times G/H$.

\begin{prop}\label{prop:2.9}
Let $X_{n,m}^t\in \mathcal{T}_C$ and 
$i(X_{n,m}^t)=[a_1:a_2:\dots,a_d]^T$.
Then it follows that $i(X_{n+1,m}^t)=i(X_{n,m+1}^t)=i(X_{n,m}^{t+1})=[a_2:a_3:\dots:a_d:a_1]^T$.
\end{prop}
\proof
Let $f_i$ be the $i$-th element of $\vect{v}$ such that $X_{n,m}^t\vect{v}=x\vect{v}$, and $f_i'$ be the $i$-th element of $\vect{v}'$ such that $X_{n+1,m}^t\vect{v}'=x\vect{v}'$. Because $X_{n+1,m}^t=(S^{m-1}L_n^tS^{-m+1})X_{n,m}^t(S^{m-1}L_n^tS^{-m+1})^{-1}$, we have $\vect{v}'=(S^{m-1}L_n^tS^{-m+1})\vect{v}$, which implies $f_i'=\left\{
\begin{array}{ll}
f_{i+1}+V_{n,m+i}^tf_i& i\neq M\\
yf_1+V_{n,m}^tf_M& i=M
\end{array}
\right.$.
From this equation and lemma \ref{lem:2.8}, we have $f_i'=
\left\{
\begin{array}{ll}
a_{i+1}k^{M-i-1}+o(k^{M-i-1})& i\neq M\\
a_{1}k^{-1}+o(k^{-1})& i=M
\end{array}
\right.$.
By multiplying $k$ to all $f_i'$ simultaneously, we obtain the desired expression. The equations about $X_{n,m+1}^t$ and $X_{n,m}^{t+1}$ are proved similarly. $\qed$

Let $\varphi$ be the injection $\mathcal{R}_C\to \mathrm{Pic}^g(C)$ introduced in \S \ref{sec:2.3}, and $\mathfrak{a}:=\mathrm{Im}\varphi$ be the image of $\varphi$. From theorem \ref{thm:2.4} and proposition \ref{prop:2.9}, we have :
\begin{thm}\label{thm:2.10}
There exist divisors $\mathcal{T}$, $\mathcal{M}_m$, $\mathcal{N}\in\mathrm{Pic}^0(C)$ and a bijection $\eta:\mathcal{T}_C\to \mathfrak{a}\times G/H$ such that: For $\eta(X_{n,m}^t)=(\mathcal{D},\vect{a})$,
\begin{gather*}
\eta(X_{n+1,m}^t)=(\mathcal{D}+\mathcal{T},\sigma(\vect{a})),\quad
\eta(X_{n,m+1}^t)=(\mathcal{D}+\mathcal{M}_m,\sigma(\vect{a})),\\
\eta(X_{n+1,m}^t)=(\mathcal{D}+\mathcal{N},\sigma(\vect{a})),\quad
\sigma([a_1:a_2:\dots:a_d]^T):=[a_2:a_3:\dots:a_d:a_1]^T.
\end{gather*}
\end{thm}

\begin{cor}
$\mathcal{T}_C$ is embedded into $\mathrm{Jac}\,C\times (\CC^\times)^{d-1}$, where $\mathrm{Jac}\,C$ is the Jacobi variety of $C$.
\end{cor}

\section{Review of tropical geometry}\label{sec:3}

We briefly review basic results of the tropical geometry. For details, see \cite{Tropical1,Tropical2}.

\subsection{ultradiscretization of functions}

Let $\RR_+^L:=(\RR_{>0})^L$ be the set of $L$-positive vectors and $T=(0,\infty)$ be the open interval of infinite length. For a topological set $A$, denote by $\mathrm{Hom}(T,A)$ the set of continuous maps from $T$ to $A$. Any continuous map $f:A\to B$ induces a mapping $f^\ast:\mathrm{Hom}(T,A)\to \mathrm{Hom}(T,B)$. 

Let $A$ be a subset of $(\CC^\times)^L$. We define the subset $\mathcal{U}^A\subset\mathrm{Hom}(T,A)$ by
\begin{equation}\label{eq:26}
\mathcal{U}^A:=\{(s_1,\dots,s_L)\in \mathrm{Hom}(T,A)\,;\, 
\textstyle\lim_{\ep\to 0^+}{\zet{\ep\log{\zet{s_i(\ep)}}}}<+\infty\}.
\end{equation}
Let $\mathrm{UD}:\mathcal{U}^A\to \RR^L$ be the map expressed as $(s_i)_i\mapsto (-\lim_{\ep\to 0^+}\ep\log{\zet{s_i(\ep)}})_i$.

\begin{defi}
A map $f:A\to B$ is ultradiscretizable if: $(i)$ $f^\ast(\mathcal{U}^A)\subset \mathcal{U}^B$, $(ii)$ There exists a map $\mathrm{UD}(f):\RR^L\to\RR^L$ such that the following diagram is commutative:
\[
\xymatrix{
\mathcal{U}^A\ar[r]^{f^\ast}\ar[d]_{\mathrm{UD}}&\mathcal{U}^B \ar[d]^{\mathrm{UD}}\\
\RR^L \ar[r]^{\mathrm{UD}(f)}& \RR^L
}.
\]
\end{defi}

We call the map $\mathrm{UD}(f)$ the \textit{ultradiscretization} of $f$. It is not easy to determine whether a given function $f$ is ultradiscretizable. Here, we introduce two simple sufficient conditions.

\begin{example}[Totally positive polynomial]
If $f$ is expressed as a totally positive polynomial, then $f$ is ultradiscretizable.
\end{example}

\begin{example}[Roots of polynomial]\label{ex:3.2}

Let $A:=(\CC^\times)^{N+1}$, $B:=\CC^\times$ and $\Phi(x)=s_Nx^N+\dots+s_1x+s_0$ with $s_i\in \mathrm{Hom}(T,\CC^\times)$. Then there exists a continuous function $x=x(\ep)$ such that $\Phi(x(\ep))=0$ for all $\ep$. Define $f_\Phi^\ast(s_0,s_1,\dots,s_N):= x$. Therefore, $f_\Phi:A\to B$ is ultradiscretizable. In fact, $\mathrm{UD}(f_\Phi)$ is a piecewise linear function of $\mathrm{UD}(s_0),\dots,\mathrm{UD}(s_N)$.
\end{example}

\subsection{tropical curves}

\subsubsection{definition}

Define the subset $\mathcal{U}:=\mathcal{U}^{\CC^\times}$ of $\mathrm{Hom}(T,\CC^\times)$ as (\ref{eq:26}). For a polynomial $\Phi=\sum_{w=(w_1,w_2)\in \ZZ^2}{c_w\, x^{w_1}y^{w_2}}\in \mathcal{U}[x^{\pm 1},y^{\pm 1}]$, denote $\mathrm{UD}(X,Y;\Phi):=\min_{w\in \ZZ^2}{[\mathrm{UD}(c_w)+w_1X+w_2Y]}$. A \textit{plane} \textit{tropical curve $\Gamma^0$ defined by} $\Phi$ is a subset of $\RR^2$ which is expressed as
\begin{align*}
\Gamma^0:=
\{(X_0,Y_0)\,\vert\,\mbox{The function } \mathrm{UD}(X,Y;\Phi)
\mbox{ is not smooth at $(X,Y)=(X_0,Y_0)$}\}.
\end{align*}

\begin{example}\label{example3.3}
Let $\ee:=e^{-1/\ep}$ and
$\Phi=y^2+y(x^3+x^2+\ee x+\ee^2)+\ee^6$.
Then, we have
$\Val(X,Y;\Phi)=\min[2Y,Y+3X,Y+2X,Y+X+1,Y+2,6]$.
The tropical curve $\Gamma^0$ is given as figure \ref{fig1a}.
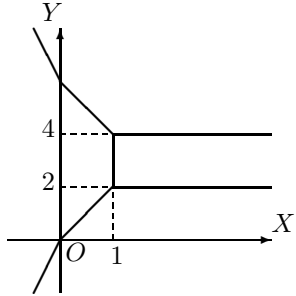
\begin{figure}[htbp]
\begin{center}
\begin{picture}(100,100)
\put(20,0){\vector(0,1){100}}
\put(0,20){\vector(1,0){100}}
%--------------------------------
\thicklines
\put(20,20){\line(-1,-2){10}}
\put(20,20){\line(0,1){60}}
\put(20,80){\line(-1,2){10}}
\put(20,80){\line(1,-1){20}}
\put(20,20){\line(1,1){20}}
\put(40,40){\line(0,1){20}}
\put(40,40){\line(1,0){60}}
\put(40,60){\line(1,0){60}}
%---------------------------------
\put(22,12){$O$}
\put(100,23){$X$}
\put(13,103){$Y$}
%--------------------------
\thinlines
\multiput(40,20)(0,4){5}{\line(0,1){2}}
\multiput(20,40)(4,0){5}{\line(1,0){2}}
\multiput(20,60)(4,0){5}{\line(1,0){2}}
\put(13,39){$2$}
\put(13,59){$4$}
\put(39,11){$1$}
\end{picture}
\end{center}
\caption{The tropical curve defined by 
$\Phi=y^2+y(x^3+x^2+\ee x+\ee^2)+\ee^6$.}
\label{fig1a}
\end{figure}
\end{example}

For $P\in \Gamma^0$, we define $\Phi^P:=\sum_{w\in \Lambda_\Phi(P)}{c_wx^{w_1}y^{w_2}}$, where $\Lambda_\Phi(P):=\{w\in \Lambda_\Phi\,\vert\,\mathrm{UD}(X_P,Y_P)=\mathrm{UD}(c_w)+w_1X_P+w_2Y_P, P=(X_P,Y_P)\}$. Let $\mathrm{mult}(P):=\sharp\{\mbox{irreducible components of } \Phi^P\!\!\in\mathcal{U}[x^{\pm 1},y^{\pm 1}]\}$.
\begin{defi}
A tropical curve with multiplicity is a non-oriented
graph $\Gamma$ with surjection $\iota:\Gamma\to \Gamma^0$
such that $\sharp\{\iota^{-1}(P)\}=\mathrm{mult}(P)$ for all
$P\in\Gamma^0$.
\end{defi}

\begin{example}
Let $\Phi$ be the polynomial in example \ref{example3.3}. Then we have, for example, $\Phi^{(1,4)}=\ee yx+\ee^2 y+\ee^6$, $\Phi^{(0,3)}=yx^3+yx^2+\ee yx=yx(x+\xi_+)(x+\xi_-)$, where $\xi_+=\ee+\ee^2+\cdots$, $\xi_-=1-\ee-\cdots$. The tropical curve with multiplicity $\Gamma$ is given as figure \ref{fig1b}.
\begin{figure}[htbp]
\begin{center}
\begin{picture}(100,100)
%--------------------------------
\thicklines
\put(20,20){\line(-1,-2){10}}
\put(20,80){\line(-1,2){10}}
\put(20,80){\line(1,-1){20}}
\put(20,20){\line(1,1){20}}
\put(40,40){\line(0,1){20}}
\put(40,40){\line(1,0){60}}
\put(40,60){\line(1,0){60}}
%---------------------------------
\qbezier(20,20)(15,50)(20,80)
\qbezier(20,20)(25,50)(20,80)
\end{picture}
\end{center}
\caption{The tropical curve with multiplicity defined by 
$\Phi$.%=y^2+y(x^3+x^2+\ee x+\ee^2)+\ee^6$.
}
\label{fig1b}
\end{figure}
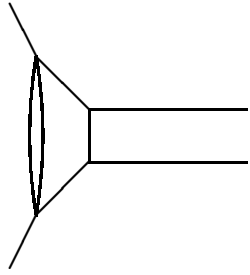
\end{example}

Note that $\iota$ is finite and locally homeomorphic without finite numbers of points. We abbreviate the pull-backed coordinates $\iota^\ast X$ and $\iota^\ast Y$ over $\Gamma$ as $X$ and $Y$.

\subsubsection{tropical integration}

Let $\iota:\Gamma\to \Gamma^0$ be a tropical curve with multiplicity. Because $\iota$ is locally homeomorphic without finite numbers of points, we can induce a metric on $\Gamma$ by pulling back the metric of $\Gamma^0$ defined by the lattice length \S5 \cite{Mikhalkin}.

Denote by $\mathcal{P}$ the free $\ZZ$-module generated by oriented paths on $\Gamma$. Let $(\cdot,\cdot):\mathcal{P}\times\mathcal{P} \to\RR$ be the \textit{tropical bilinear form} \cite{Mikhalkin}, which is defined by $(\gamma_1,\gamma_2):=(\pm1)\cdot \kakko{\gamma_1\cap\gamma_2}$.  Fix a basis $\beta_1,\dots,\beta_g$ of $H_1(\Gamma;\ZZ)$. Then, we obtain the \textit{period matrix} of $\Gamma$, which is a $g\times g$ matrix $B_\Gamma=\left((\beta_i,\beta_j)\right)_{i,j=1}^g$. For a fixed point $Q_0\in\Gamma$, we define the \textit{tropical Abel-Jacobi mapping} $F_{Q_0}:\Gamma\to\RR^g/B_\Gamma \ZZ^g$ by 
\[
F_{Q_0}(P):=\left(\phantom{\sum}\hspace{-15pt}(\gamma_{Q_0\to P},\beta_i)\right)_{i=1}^g
\qquad (\mathrm{mod }\ B_\Gamma\ZZ^g),
\]
where $\gamma_{Q_0\to P}\in\mathcal{P}$ is a path from $Q_0$ to $P$. We call the variety $J(\Gamma):=\RR^g/B_\Gamma \ZZ^g$ the \textit{tropical Jacobi variety}. The mapping $F_{Q_0}$ can be linearly extended to $\mathrm{Div}\,\Gamma=\bigoplus_{p\in\Gamma}\ZZ\cdot p$.

\subsection{ultradiscrete limit of Abelian integrals}\label{sec:3.3}

In this section, we review the theorem in the paper \cite{Iwao}, which explains a certain direct relation between Abelian integrals and tropical integrals. For a polynomial $\Phi(x,y)\in \mathcal{U}[x^{\pm 1},y^{\pm 1}]$, there exists a holomorphic family of curves $\pi:V\to T$ such that $\pi^{-1}(\ep)$ is an algebraic curve defined by $\Phi\vert_\ep\in \CC[x^{\pm 1},y^{\pm 1}]$. Denote by $g$ the genus of these curves.

Let $\alpha_i:A\to T$ and $\beta_i:B\to T$ ($i=1,\dots,g$) be a holomorphic (oriented) subfamily of $\pi$ such that $\alpha_i^{-1}(\ep)$ and $\beta_i^{-1}(\ep)$ are symplectic basis of $H_1(\pi^{-1}(\ep);\ZZ)$. We call such $(\alpha_i,\beta_i)$ a \textit{symplectic basis of } $\pi$.

The following is the main result of the paper \cite{Iwao}.

\begin{thm}[\cite{Iwao}]\label{thm:3.5}
There exists a canonical correspondence
\[
\mathfrak{X}:
\{(\alpha_i,\beta_i)_i\,\vert\,\mbox{symplectic basis of $\pi:V\to T$}\}\to
\{(\beta_i)_i\,\vert\,\mbox{basis of } H_1(\Gamma;\ZZ)\}
\]
and holomorphic $1$-forms $\omega_i$ $(i=1,\dots,g)$ over $V$ such that 
\[
\int_{\alpha_i^{-1}(\ep)}\omega_j=\delta_{i,j},\quad
-2\pi i\lim_{\ep\to 0}\left(\ep\int_{\beta_i^{-1}(\ep)}\omega_j\right)=B_{i,j},
\]
where $B_{i,j}$ is the $(i,j)$-component of $B_\Gamma$ defined by the basis $\mathfrak{X}(\alpha_i,\beta_i)$. 
$\hfill\blacksquare$
\end{thm}

Let $P\in\Gamma$. Because the field $\mathcal{U}$ is algebraically closed, we can find a holomorphic section $p:T\to V$ of $\pi$
such that $\mathrm{UD}(x(p(\ep)))=X(P)$, $\mathrm{UD}(y(p(\ep)))=Y(P)$. In this case, we denote `$p\ud P$'.

\begin{thm}\label{thm:3.6}
Let $p,q$ be two sections $T\to V$.
Assume there exist two points 
$P,Q\in\Gamma$ such that
$p\ud P$, $q\ud Q$. 
Then, the following relations
are held
\begin{gather*}
-2\pi i\lim_{\ep\to 0^+}{\ep\cdot f_{q(\ep)}(p(\ep))}=F_{Q}(P)\ \ \in 
J(\Gamma).\ \ \ \hfill\qed
\end{gather*}
\end{thm}

\section{Application for ultradiscrete systems}\label{sec4}

From the results in \S\ref{sec:2} and \S\ref{sec:3}, we obtain the tropical analog of the linearizing theorem for the ultradiscrete integrable systems. Let $J$ be the quotient space $J:=(\CC^g\times \mathrm{GL}(g,\CC))/\sim$, where $(\vect{v},B)\sim (\vect{v}',B') \ \Leftrightarrow\ B=B'\mbox{ and } \vect{v}-\vect{v}'\in \ZZ^g+B\ZZ^g$. The analytical set $J$ is regarded as a set of \textit{all} Jacobi varieties. Any element of $J$ is expressed as $(\vect{v}\ \mathrm{mod}\,(\ZZ^g+B\ZZ^g),B)$.

\subsection{ultradiscretization}

\subsubsection{family of isolevel sets}

For an $\beta\in\mathrm{Hom}(T,\CC^\times)$, define
\[
\mathcal{V}(\beta):=
\left\{
\left.
(V_{n,m})_{n=1,m=1}^{N,M}\in \mathrm{Hom}(T,\CC^{N\times M})\,\right\vert \,
\begin{array}{ll}
(i)\, \beta=\prod_m{V_{n,m}} \mbox{ for all } n,\\
(ii)\,
\textstyle\lim_{\ep\to 0^+}{\zet{\ep\log{\zet{V_{n,m}(\ep)}}}}<+\infty
\!\end{array}
\right\}.
\]
As in \S\ref{sec3.1}, any $\{V_{n,m}\}_{n,m}\in\mathcal{V}(\alpha)$ is identified with the matrix function $X_n:=L_{n+N-1}\cdots L_{n+1}L_n$, where $L_n=\mathrm{diag}(V_{n,1},\dots,V_{n,M})+S$. Let $\pi:V\to T$ be the holomorphic family of curves defined by $\Phi=\det{(X_n-x\cdot\mathrm{id.})}$. Consider the correspondence: $\mathrm{SC}^\ast: \mathcal{V}(\alpha)\to \{\mbox{hol. families over } T\};\ \{V_{n,m}\}_{n,m}\mapsto \{\pi:V\to T\}$. For a holomorphic family $\pi:V\to T$, denote $\mathcal{T}_\pi:=(\mathrm{SC}^\ast)^{-1}(\pi)$. This $\mathcal{T}_\pi$ is regarded as a holomorphic family of isolevel sets.

For any $\ep\in T$, the isolevel set $\mathcal{T}_\pi\vert_\ep$ can be embedded into $\mathrm{Jac}(\pi^{-1}(\ep))\times (\CC^\times)^{d-1}$ (theorem \ref{thm:2.10}). Combining these maps, we obtain the inclusion $\eta^\ast:\mathcal{T}_\pi\to \mathrm{Hom}(T,J\times (\CC^\times)^{d-1})$ induced from $\eta$.

\subsubsection{ultradiscretization of $\eta$}

We are interested in the real positive part of the isolevel set: $\mathcal{V}(\alpha)^+:=\mathcal{V}(\alpha)\cap \mathrm{Hom}(T,\RR_+^{N\times M})$. Let $\mathcal{T}_\pi^+:=\mathcal{T}_\pi\cap \mathrm{Hom}(T,\RR_+^{N\times M})$.
Denote by $\eta^\ast_+:\mathcal{T}_\pi^+\to \mathrm{Hom}(T,J\times (\CC^\times)^{d-1})$ 
the restriction of $\eta^\ast$ to $\mathcal{T}_\pi^+$. Let $\mathcal{L}:\mathrm{Hom}(T,J\times (\CC^\times)^{d-1})\to J(\Gamma)\times \RR^{d-1}$ be the mapping defined by
\[
((\vect{v}(\ep)\ \mathrm{mod}\,(\ZZ^g+B(\ep)\ZZ^g),B(\ep)),\vect{w}(\ep))\mapsto {(-2\pi i\lim_{\ep\to 0^+}\ep\vect{v}(\ep),
-\lim_{\ep\to 0^+}\ep\log \vect{w}(\ep))}.
\]
By theorem \ref{thm:3.5}, this mapping is well-defined.

Our aim in this section is to prove the following theorem:
\begin{thm}\label{thm:4.1}
There 
exists a mapping $\mathrm{UD}(\eta)$ such that the following 
diagram is commutative:
\[
\xymatrix{
\mathcal{T}^+_\pi \ar[r]^{\hspace{-40pt}\eta_+^\ast}\ar[d]^{\mathrm{UD}}
& \mathrm{Hom}(T,J\times (\CC^\times)^{d-1})
\ar[d]^{\mathcal{L}}\\
\RR^{N\times M} \ar[r]^{\mathrm{UD}(\eta)}& J(\Gamma)\times \RR^{d-1}
},
\]
where $\Gamma$ is the tropical curve associated with $\Phi$.
\end{thm}

First, we refer to the result proved by Mada, Idzumi and Tokihiro \cite{Tokihiro3}:
\begin{thm}\label{thm:4.2}
$(i)$ Let $L_n=\mathrm{diag}(V_{n,1},\dots,V_{n,M})+S$ and $X:=L_N\cdots L_2L_1$. Then, all the coefficients of $\Phi(x,y)=\det (X(y)-xE)$ are expressed as $\pm($totally positive polynomial in $V_{n,m}$ $(n=1,\dots,N,m=1,\dots,M))$.\\
$(ii)$\footnote{In the paper \cite{Tokihiro3}, only the proof of the part (i) is given. However, the part (ii) is also proved by the exactly same argument.} Coefficients of any $(i,j)$-minor $\Delta_{i,j}(x,y)$ of $(X(y)-xE)$ are expressed as $\pm($totally positive polynomial in $V_{n,m})$. $\hfill\blacksquare$
\end{thm}
{\bf Proof of theorem \ref{thm:4.1}.}\ 
We prove the theorem by constructing
$\mathrm{UD}(\eta)_1:\RR^{N\times M}\to J(\Gamma)$
and $\mathrm{UD}(\eta)_2:\RR^{N\times M}\to \RR^{d-1}$
respectively.

(I) Constructing $\mathrm{UD}_1$:
For a holomorphic family $\pi:V\to T$ of analytic curves of genus $g$, we denote by $\mathrm{Div}^g(\pi) :V_g\to T$ the holomorphic family of divisors of degree $g$. A fiber of $\mathrm{Div}^g(\pi)$ is of the form $\{\mathrm{Div}^g(\pi)\}^{-1}(\ep)=\mathrm{Div}^g(\pi^{-1}(\ep))$. By theorem \ref{thm:2.3}, we have the mapping $\tilde{\varphi}_C:\mathcal{R}_C=\mathcal{T}_C/\sim\ \to \mathrm{Div}^g(C)$ for any spectral curve $C$. Therefore, this mapping $\tilde{\varphi}_C$ induces a new mapping $\tilde{\varphi}^\ast:\mathcal{T}_\pi\to\mathrm{Div}^g(\pi)$.

Let $v=(V_{m,n}(\ep))_{m,n}$ be an element of $\mathcal{T}_\pi$. Denote $X_v=L_N\cdots L_2L_1$, $L_n=\mathrm{diag}(V_{n,1},\dots,V_{n,M})+S$. By theorem \ref{thm:2.3} (ii), $x,y$ coordinates of $g$ points which belongs to $\tilde\varphi^\ast(X_v)(\ep)$ are expressed as roots of polynomials defined by minors of $(X_v-xE)$. Due to theorem \ref{thm:4.2}, all the coefficients of these polynomials are of the form $\pm$(totally positive polynomials in $V_{n,m}$). Therefore, the limits $\{-\lim_{\ep\to 0^+}{\ep\log x_i(\ep)}\}_{i=1}^g$, $\{-\lim_{\ep\to 0^+}{\ep\log y_i(\ep)}\}_{i=1}^g,$ $(x_i,y_i=\{x,y\mbox{ coordinates of a point } p_i(\ep) \mbox{ where }\tilde\varphi^\ast(X_v)(\ep)=p_1+\dots+p_g  \})$ depend only on $\mathrm{UD}(X(\ep))$. Let $P_i$ be a point in $\Gamma$ such that $p_i\ud P_i$ (\S \ref{sec:3.3}) and $p_0:T\to V$ be an arbitrary section and $p_0\ud P_0\in \Gamma$. By theorem \ref{thm:3.6}, it is sufficient to define $\mathrm{UD}(\eta)_1(X):=\sum_{i=1}^gF_{P_0}(P_i)$, where $X\in\RR^{M\times N}$.

(II) Constructing $\mathrm{UD}(\eta)_2$:
We construct $\mathrm{UD}(\eta)_2$ by ultradiscretizing the mapping $i:\mathcal{T}_C\to (\CC^\times)^{d-1}$ introduced in \S \ref{sec:2.4}. By the proof of lemma \ref{lem:2.8}, the image $i(X)$ is expressed as $[a_1:\dots:a_d]$, where any $a_i$ is a root of some polynomials of which coefficients are components of $X$. Therefore, the expression $\mathrm{UD}(\eta)_2(X):=(-\lim_{\ep\to 0^+}{\ep\log \zet{a_i}})_i$, $X=\mathrm{UD}(X(\ep))$ is well-defined by example \ref{ex:3.2}. $\qed$

By the analogy of the discrete case \cite{Moerbeke}, we can make a conjecture concerning the ultradiscretized 
$\tilde\eta$.
\begin{conj}\label{conj}
The mapping $\mathrm{UD}(\eta)$ is injective.
\end{conj}

\subsection{fundamental cycles of 2dBBS}\label{sec4.2}

Now we consider the 2dBBS (\S \ref{sec1}). Let $\{W_{n,m}^t,Q_{n,m}^t\}_{n,m}^t$ be a set of real numbers which satisfies the 2dBBS (\ref{eq:6}). The \textit{fundamental cycle of 2dBBS} is the minimum positive integer $F$ such that 
\[
\{W_{n,m}^F,Q_{n,m}^F\}_{n,m}=\{W_{n,m}^0,Q_{n,m}^0\}_{n,m}.
\]

Let $V_{n,m}^0=V_{n,m}^0(\ep)$ be a real function satisfying $\mathrm{UD}(V_{n,m}^0)=W_{n,m}^0$. Using the procedure in \S \ref{sec3.1}, we obtain the matrix function $X(y)(\ep)\in\mathrm{Hom}(T,\mathrm{Mat}(M,\CC[y]))$ and therefore the family of spectral curves $\pi:V\to T$. Denote by $\Gamma$ the tropical curve defined by $X(y)(\ep)$. %According to theorem \ref{thm:4.1}, we can send the dynamics of 2dBBS to $\mathrm{UD}(\eta):\RR^{N\times M}\to J(\Gamma)\times \RR^{d-1}$.

Here we recall proposition \ref{prop:2.6}. We denote by $p_0(\ep)$, $p_1(\ep)$, $p_2(\ep)$, $p^{(m)}_3(\ep)$ the points in $C_\ep=\pi^{-1}(\ep)$ as in proposition \ref{prop:2.6}
\begin{prop}\label{prop:4.4}
Let $P_0,P_1,P_2,P_3^{(m)}\in\Gamma$ be the tropicalization of $p_0$, $p_1$, $p_2$, $p_3^{(m)}$. $(p_i^{(m)}\ud P^{(m)}_i)$. Then, we have $P_0:(X,Y)=(-\infty,-\infty)$, $P_1:(X,Y)=(G,A)$, $P_2:(X,Y)=(+\infty,B)$, $P_3^{(m)}:(X,Y)=(H_m,+\infty)$, where $A=\sum_{m=1}^M{Q_{n,m}^0}$, $B=\sum_{m=1}^MW_{n,m}^0$, $H_m=\sum_{n=1}^N{W_{n,m}^0}$ and $G$ is the maximum number such that $(G,A)\in\Gamma$.
\end{prop}
\proof $\val(\alpha)=A$, $\val(\beta)=B$ and $\val(\gamma_m)=H_m$ are obtained by definition. The $G$ is obtained from $G=\val{(\kappa)}$, where $\kappa$ is the complex number in lemma \ref{lem:2.5} (ii).
$\qed$

Using $\Gamma$, we can calculate the tropical period matrix $B_{\Gamma}$ and vectors $\vec{\mathcal{T}}:=F_{P_1}(P_0)$, $\vec{\mathcal{N}}:=F_{P_2}(P_0)$, $\vec{\mathcal{M}}^{(m)}:=F_{P_3^{(m)}}(P_0)\in J(\Gamma)$ by combinatorial procedures.

\begin{prop}\label{prop:4.5}
Let $F''$ be the minimum positive integer such that $F''\cdot \vec{\mathcal{T}}\equiv 0$ and $F':=\mathrm{l.c.m.}(F'',d)$. The fundamental cycle $F$ is a multiple of $F'$. If conjecture \ref{conj} is true, then $F=F'$.
\end{prop}
\proof
This proposition is the consequence of theorems \ref{thm:2.10}, \ref{thm:3.6}, \ref{thm:4.1}.
$\qed$

Note that the condition $F\cdot \vec{\mathcal{T}}\equiv 0\in J(\Gamma)$ is equivalent to $F\cdot B_{\Gamma}^{-1}\vec{\mathcal{T}}\in \ZZ^g$.

\subsubsection{example I}

%In the rest of the paper, $\ee$ denotes the continuous function
Let $\ee:=e^{-1/\ep}\in \mathrm{Hom}(T,\RR_+)$. Consider the following example:
\begin{verbatim}
             1|..1      .|1..      .|.1.      1|..1
             .|.1.      1|..1      .|1..      .|.1.
             .|1..      .|.1.      1|..1      .|1..
              t=0        t=1        t=2        t=3
\end{verbatim}
and $A=B=1$, $M=N=3$, $d=3$. From the picture, we have:
\begin{align*}
X_1^0=L_3^0L_2^0L_1^0%\\
=(\mathrm{diag}(\ee^1,\ee^0,\ee^0)+S)
(\mathrm{diag}(\ee^0,\ee^1,\ee^0)+S)(\mathrm{diag}(\ee^0,\ee^0,\ee^1)+S).
\end{align*}
Its characteristic polynomial $\Phi(x,y)$ satisfies 
\begin{align*}
\Phi(x,y)=\det(X_1^0(y)-xE)
=-x^3+x^2(3y+3\ee)-x(3y^2-21\ee y+3\ee^2)+(y+\ee)^3.
\end{align*}
Therefore, the tropical curve $\Gamma$ defined by $\Phi$ satisfies 
$B_\Gamma=0$, which implies $J(\Gamma)=\{0\}$. 
By proposition \ref{prop:4.5}, the fundamental cycle 
is a multiple of $d=3$.
(In fact, the fundamental cycle is $3$.)

\subsubsection{example II}

Start with the following initial value:
\begin{verbatim}
         .|..11  .|..2.  .|1..1  1|...2        .|..11
         .|.11.  .|.2..  .|..11  .|..2.   ...  .|.11.
         1|21.1  1|2..2  1|121.  .|22..        1|21.1
           t=0     t=1     t=2     t=3           t=8
\end{verbatim}
and $A=1,B=2$, $M=3,N=4$, $d=1$. Form the picture, we have: %$X_1^0(=L_4^0L_3^0L_2^0L_1^0)$: 
\begin{align*}
& X_1^0=(\mathrm{diag}(\ee^2,1,1)+S)(\mathrm{diag}(\ee,\ee,1)+S)
(\mathrm{diag}(1,\ee,\ee)+S)(\mathrm{diag}(\ee,1,\ee)+S).
\end{align*}
Then $\Phi(x,y)=-x^3+x^2(5y+2\ee^2)-x(y^2-7\ee^2y+\ee^4)+(y+\ee^2)^4+(\mathrm{small})$. Therefore, the tropical curve $\Gamma$ defined by $\Phi$ is as figure \ref{fig:5}:

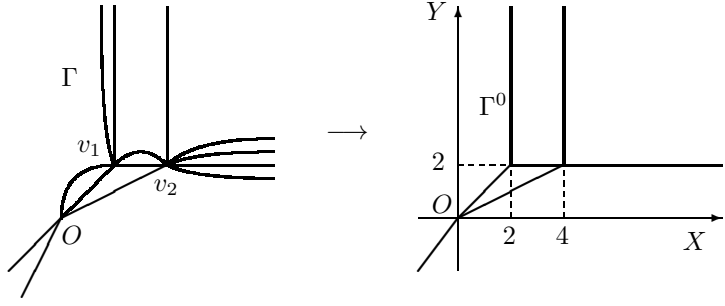
\begin{figure}[htbp]
\begin{center}
\begin{picture}(270,100)
\put(170,0){\vector(0,1){100}}
\put(155,20){\vector(1,0){115}}
%--------
\thicklines
\put(155,0){\line(3,4){15}}
\put(170,20){\line(2,1){40}}
\put(170,20){\line(1,1){20}}
\put(210,40){\line(1,0){60}}
\put(190,40){\line(1,0){20}}
\put(190,40){\line(0,1){60}}
\put(210,40){\line(0,1){60}}
%-----------
\thinlines
\multiput(170,40)(4,0){5}{\line(1,0){2}}
\multiput(190,20)(0,4){5}{\line(0,1){2}}
\multiput(210,20)(0,4){5}{\line(0,1){2}}
\put(160,22){$O$}
\put(160,38){$2$}
\put(187,10){$2$}
\put(207,10){$4$}
\put(255,8){$X$}
\put(158,95){$Y$}
\put(178,59){$\Gamma^0$}
%------------------------------------------
\thicklines
\put(20,20){\line(-1,-2){15}}
\put(20,20){\line(-1,-1){20}}
\put(20,20){\line(2,1){40}}
%---
\qbezier(20,20)(30,30)(40,40)
\qbezier(20,20)(20,40)(40,40)
%---
\qbezier(40,40)(50,40)(60,40)
\qbezier(40,40)(50,50)(60,40)
%---
\put(60,40){\line(0,1){60}}
%---
\qbezier(40,40)(35,50)(35,100)
\qbezier(40,40)(40,50)(40,100)
%\qbezier(40,40)(45,50)(45,100)
%---
\qbezier(60,40)(70,50)(100,50)
\qbezier(60,40)(70,45)(100,45)
\qbezier(60,40)(70,40)(100,40)
\qbezier(60,40)(70,35)(100,35)
%----
\put(20,70){$\Gamma$}
\put(20,10){$O$}
\put(26,45){$v_1$}
\put(55,30){$v_2$}
%---------------------------
\put(120,50){$\longrightarrow$}
\end{picture}
\end{center}
\caption{Tropical curve with multiplicity $\Gamma\to \Gamma^0$
in Example II.}
\label{fig:5}
\end{figure}

Let $O=(0,0)$, $v_1=(2,2)$ and $v_2=(4,2)$. Denote two edges connecting $O$ and $v_1$ by $e_1$ and $e_2$, and two edges connecting $v_1$ and $v_2$ by $e_3$ and $e_4$. Define closed paths $\beta_1,\beta_2,\beta_3$ on $\Gamma$ as follows:
\begin{gather*}
\beta_1:v_1 \stackrel{e_1}{\to} O \stackrel{e_2}{\to} v_1,\quad
\beta_2:v_2 \stackrel{e_3}{\to} v_1 \stackrel{e_4}{\to} v_2,\quad
\beta_3:O \to v_2 \stackrel{e_4}{\to} v_1 \stackrel{e_2}{\to} O.
\end{gather*}
Therefore, the tropical period matrix is 
$
B_\Gamma=
\left(
\begin{array}{@{\,}ccc@{\,}}
	4 & 0 & -2 \\
	0 & 4 & -2 \\
	-2 & -2 & 6
\end{array}\right)
$.
By proposition \ref{prop:4.4}, we have $P_0=(-\infty,-\infty)$, $P_1=(2,1)$, $P_2=(+\infty,2)$, $P_3^{(1)}=(4,+\infty)$ and $P_3^{(2)}=P_3^{(3)}=(2,+\infty)$. Then, $\vec{\mathcal{T}}=(0,0,-1)^T,\ \vec{\mathcal{N}}=(0,0,-2)^T,\ \vec{\mathcal{M}}^{(1)}=(2,2,0)^T,\ \vec{\mathcal{M}}^{(2)}=\vec{\mathcal{M}}^{(3)}=(2,0,0)^T$. It follows that $(B_\Gamma)^{-1}\vect{v}_\mathcal{T}=(-1/8,-1/8,-1/4)^T$, which implies that the fundamental cycle is a multiple of $8$. (If fact, the fundamental cycle is $8$).
We note the relations $4\vec{\mathcal{N}}=B_\Gamma (-1,-1,-2)^T\in B_\Gamma \ZZ^3$ and $\vect{v}_{\mathcal{M}}^{(1)}+\vect{v}_{\mathcal{M}}^{(2)}+\vect{v}_{\mathcal{M}}^{(3)}=B_\Gamma (2,1,1)^T\in B_\Gamma\ZZ^3$, which reflect the condition $X_{n+4,m}^t=X_{n,m+3}^t=X_{n,m}^t$.

\subsection*{Acknowledgment}

The author is very grateful to Professor Tetsuji Tokihiro and Professor Ralph
Willox for helpful comments on this paper.
Akane Nakamura gave the author a hint 
for the proof of proposition \ref{prop:2.2}.
The author appreciate the dedicated support by
Kayo Ejiri.
This work was supported by KAKENHI 23-1939.

\appendix

\section{Proof of lemmas}

\subsection{Proof of lemma \ref{lem:2.8}}

Let $C$ be the spectral curve defined in \S\ref{sec3.1}, $p_0$ be the unique point contained in $C\setminus C_0$ and $k$ be the local coordinate around $p_0$. Let $S_k:=(\delta_{i+1,j})+k^{-M}E_{M,1}$. At $p_0$, the equation $X\vect{v}=x\vect{v}$ implies $k^N(L_N+S_k)\cdots(L_2+S_k)(L_1+S_k)\vect{v}=(1+O(k))\vect{v}$.
\begin{lemma}[lemma \ref{lem:2.8}]
Up to multiplication by a constant, the $i$-th component $f_i$ of $\vect{v}$ has the following expression near $p_0$:
\[
f_i=a_ik^{M-i}+o(k^{M-i}),\qquad a_i\neq 0.
\]
Moreover, we have $a_{i+d}=a_i$ for all $i$.
\end{lemma}
\proof
Let $d=\mathrm{g.c.d.}(N,M)$, $N=dN_1$ and $M=dM_1$. Denote $Y_{p,q}:=(y_{pd+i,qd+j})_{i,j=1}^d$ for a square matrix $Y=(y_{i,j})_{i,j=1}^M$ of size $M$. Let $K=(\kappa_{i,j})$ be the diagonal matrix such that $(p-1)d<i\leq pd\ \Rightarrow\ \kappa_{i,i}=k^{M-pd}$, $p=1,2,\dots,M_1$. Putting $Y:=k^NK^{-1}(L_N+S_k)\cdots(L_2+S_k)(L_1+S_k)K$ and $\vect{w}:=K^{-1}\vect{v}$, we obtain $Y\vect{w}=(1+c_1k+o(k))\vect{w}$. ($c_1$ is a non-zero constant). Because the $(i,j)$-element $x_{i,j}$ of $(L_N+S_k)\cdots(L_2+S_k)(L_1+S_k)$ satisfies $x_{i,j}\in \CC[k]$ $(i\leq j)$, $x_{i,j}\in k^{-M}\CC[k]$ $(i> j)$ and $x_{i,j+N}=1$ ($x_{i,j+M}\equiv x_{i,j}k^{-M}$), the $(p,q)$-th block $Y_{p,q}$ satisfies
\begin{gather*}
Y_{p,q}=\left\{
\begin{array}{ll}
E+(\mbox{strictly lower triangle mat.})+O(k^d),&
q-p\equiv N_1 (\mbox{mod } M_1),\\
O(k^d), & \mbox{otherwise}.
\end{array}
\right.
\end{gather*}
Because $\mathrm{g.c.d.}(N_1,M_1)=1$, the vector $\vect{w}$ must be expressed as:
\begin{equation}\label{eq:a.2}
\vect{w}=(\overbrace{0,0,\dots,0,1}^d,\overbrace{0,0,\dots,0,1}^d,
\dots,\overbrace{0,0,\dots,0,1}^d)^T+O(k^d).
\end{equation}
On the other hand, for the matrix $Z=Y^{M_1}$,
the $(p,q)$-th block $Z_{p,q}$ satisfies
\begin{equation}\label{eq:a.3}
 Z_{p,q}=\left\{
\begin{array}{ll}
Y_{p,p+N_1}Y_{p+N_1,p+2N_1}Y_{p+2N_1,p+3N_1}\cdots Y_{p+(M_1-1)N_1,p} & p=q,\\
O(k^d) & p\neq q,
\end{array}
\right.
\end{equation}
where $Y_{p+{M_1},q}\equiv Y_{p,q+M_1}\equiv Y_{p,q}$. Then the equation $Y\vect{w}=(1+c_1k+o(k))\vect{w}$ induces
\begin{equation}\label{eq:a.4}
Z_{p,p}\vect{w}_p\equiv (1+M_1c_1k+o(k))\vect{w}_p\ \ 
(\mathrm{mod}\,k^d),
\qquad  p=1,2,\dots,M_1,
\end{equation}
where $\vect{w}=(g_i)_{i=1}^M$, $\vect{w}_p=(g_{d(p-1)+i})_{i=1}^d$. From this, there exist non-zero complex numbers $a_1^{(p)},\dots,a_d^{(p)}$ such that
\begin{equation}\label{eq:a.5}
\vect{w}_p=(a_1^{(p)}k^{d-1},a_2^{(p)}k^{d-2},\dots,
a_d^{(p)})+(\mbox{smaller when $k\to 0$}).
\end{equation}
From (\ref{eq:a.2}, \ref{eq:a.5}), we obtain $g_{d(p-1)+i}=a_i^{(p)}k^{d-i}+o(k^{d-i})$ and $a_d^{(1)}=\cdots=a_d^{(M_1)}$. Using the equation $\vect{v}=K\vect{w}$, we rewrite these equations as  $f_i=a_ik^{M-i}+o(k^{M-i})$ and $a_d=a_{2d}=\cdots=a_{M-d}$.

We prove the equation $a_{i+d}=a_i$ through the following two steps:\\
{\bf Step 1.}\ Let $z_{i,j}^{(p)}$ be the $(i,j)$-component of the matrix $Z_{p,p}$. We will prove $z_{i+1,i}^{(1)}\equiv z_{i+1,i}^{(2)}\equiv\cdots\equiv z_{i+1,i}^{(M_1)}$ $(\mathrm{mod} k^d)$. Let $Y^-_{p}$ be a lower triangle matrix with diagonal elements $1$ such that $Y_{p,p+N_1}\equiv Y^-_{p}$ $(\mathrm{mod}\ k^d)$. Let $T:=(\delta_{i,j+1})_{i,j=1}^d$. Then the matrix $Y^-_{p}$ is decomposed uniquely as $Y^-_{p}=E+\sum_{i=1}^{d-1}{U^{(p)}_i T^i}$, $(U^{(p)}_i:\mbox{diagonal})$. Therefore, we have $Y^-_{p_1}Y^-_{p_2}\cdots Y^-_{p_l}=I+(U_1^{(p_1)}+U_1^{(p_2)}+\cdots+U_1^{(p_l)})T+\cdots$, which implies that the $(i+1,i)$-component of the matrix $Y^-_{p_1}Y^-_{p_2}\cdots Y^-_{p_l}$ does not depend on the order of $p_i$. Because $\{p,p+N_1,\dots,p+(M_1-1)N_1\}\equiv \{1,2,\dots,M_1\}$ $(\mathrm{mod}\ M_1)$ for all $p$, we conclude the claim of this step by (\ref{eq:a.3}).\\
{\bf Step 2.}\ Prove $z_{i+1,i}^{(p)}a_i^{(p)}\equiv M_1c_1a_{i+1}^{(p)}$
$(\mathrm{mod}\ k^d)$. This is a direct consequence of (\ref{eq:a.4}).

From these two steps, we have $a_i^{(p)}=a_i^{(p+1)}$ which is equivalent to $a_{i}=a_{i+d}$. $\qed$

\subsection{Proofs of propositions \ref{prop:2.1}, \ref{prop:2.2}.}

Let $I_n^t=\mathrm{diag}(I_{n,1}^t,\dots,I_{n,M}^t)$, $V_n^t=\mathrm{diag}(V_{n,1}^t,\dots,V_{n,M}^t)$. Then the equation $X_{n}^{t}=L_{n+N-1}^t\cdots L_{n+1}^tL_n^t$
is rewritten as 
\[
X_n^t=Y_0^t+SY_1^t+\dots+S^{N-1}Y_{N-1}^t+S^N
=Z_0^t+Z_1^tS+\dots+Z_{N-1}^tS^{N-1}+S^N,
\]
where $Y_n^t,Z_n^t$ ($n=0,1,\dots,N-1$) are diagonal matrices. We note that the matrices $Y_0^t,\dots,Y_{N-1}^t$ are independent as functions of $\{V_{n,m}^t\}_{k,m}$.

\begin{prop}[proposition \ref{prop:2.1}]
For given $X^t_n\in\mathcal{W}(\beta)$, there exist $M$ pairs $(R^{(1)},X^{(1)}),\cdots$, $(R^{(M)},X^{(M)})$ such that $(i)$ $R^{(i)}=\mathrm{diag}(I^{(i)}_{1},\dots,I^{(i)}_M)+S$, $(ii)$ $\alpha=\prod_{m=1}^M{I_m^{(i)}}$, $(iii)$ $X^{(i)}\in\mathcal{W}(\beta)$, $(iv)$ $X^{(i)}R^{(i)}=R^{(i)}X_n^t$.
\end{prop}

\proof
It is sufficient to prove for a generic $X_n^t$. In this proof, we denote $I=I_m^{(i)}$, $Y_n=Y_n^t$, $Z_n=Z_n^{t+1}$ for simplicity. Then the equation $(\ref{eq:12})$ is rewritten as 
\begin{align*}
(Z_0+Z_1S+\cdots+Z_{N-1}S^{N-1}+S^N)(I+S)=(I+S)(Y_0+\cdots+S^{N-1}Y_{N-1}+S^N),
\end{align*}
which implies $Z_kS^kI+Z_{k-1}S^k=IS^kY_k+S^kY_{k-1}$, $(k=0,1,\dots,N)$, $Z_{-1}=Y_{-1}=0$, $Z_N=Y_N=E$.
Using these relations recursively and deleting $Z_k$, we have 
\[
0=Y_0-Z_0=[Y_0-SY_0S^{-1}]-I(SY_1S^{-1})+(SIS^{-1})Z_1=\cdots=\sum_{k=0}^N{(-1)^k(\Theta_k-S\Theta_kS^{-1})}, 
\]
where $\Theta_k=I(SIS^{-1})(S^2IS^{-2})\cdots(S^{k-1}IS^{-k+1})(S^kY_kS^{-k})$. This implies that the diagonal matrix $\mathcal{M}:=\sum_{k=0}^{M}(-1)^k\Theta_k$ must satisfy $\mathcal{M}=S\mathcal{M}S^{-1}$, which is equivalent to $\mathcal{M}=\kappa E$ for a certain constant $\kappa$. Let $I=-\mathrm{diag}(x_1,\dots,x_M)$ and $S^iY_iS^{-i}=\mathrm{diag}(c_{i,1},c_{i,2},\dots,c_{i,M})$. Then $\mathcal{M}=\kappa E$ is rewritten as
\begin{equation}\label{eq:a.7}
\textstyle \kappa=c_{0,m}+\sum_{i=1}^{N-1}{c_{i,m}x_mx_{m+1}\cdots x_{m+i-1}}+x_mx_{m+1}\cdots x_{m+N-1}
\end{equation}
for all $m\in\{1,2,\dots,M\}$ ($x_{m+M}\equiv x_m$). Let $\mu_i$ be a new variable defined by $x_i=\frac{\mu_{i+1}}{\mu_i}$. Then, we have $\frac{\mu_{M+i}}{\mu_i}=x_1x_2\cdots x_M=(-1)^M\alpha$ for any $i$. Therefore, (\ref{eq:a.7}) is rewritten as $\kappa=\frac{\sum_{i=0}^{N-1}{c_{i,m}\mu_{m+i}}+\mu_{m+N}}{\mu_m}$, $\forall m$. Putting $S_\alpha:=S\vert_{y=(-1)^M\alpha}$ and $\vect{\mu}:=(\mu_1,\dots,\mu_M)^T$, we obtain $(Y_0+S_\alpha Y_1+S_\alpha^2 Y_2+\cdots+S_\alpha^{N-1} Y_{N-1}+S_\alpha^N)\vect{\mu}=\kappa\vect{\mu}$, which implies 
\begin{equation}\label{eq:a.8}
(X_n^t\vert_{y=(-1)^M\alpha})\vect{\mu}=\kappa\vect{\mu}.
\end{equation}
Because $X_n^t$ is generic, there exist $M$ eigenvectors $\vect{\mu}$ with non-zero elements.
$\qed$ 

To prove proposition \ref{prop:2.2}, we review the well-known result on linear algebra: %proved by Perron and Frobenius:
\begin{lemma}[the Perron-Frobenius theorem]\label{lem:a.3}
Let $X$ be a positive matrix. 
Then, $(i)$ $X$ has positive eigenvalues, $(ii)$ the maximum eigenvalue $\kappa_{\mathrm{max}}$ in absolute value is positive and simple, $(iii)$ any positive eigenvector of $X$ belongs to $\kappa_{\mathrm{max}}$. $\qed$
\end{lemma}

\begin{prop}[Proposition \ref{prop:2.2}]
Let 
\[
X_n^t=L_{n+N-1}\cdots L_{n+1}L_n\in\mathcal{W}(\beta),\quad L_n^t=\mathrm{diag}(V_{n,1}^t,\dots,V_{n,M}^t)+S \quad\mbox{such that}\quad V_{n,m}^t>0.
\] 
Then, there uniquely exist two matrices $X_n^{t+1}$, $R_n^t$ such that $(i)$ $R_n^t=\mathrm{diag}(I^t_{1},\dots,I^t_M)+S$, $(ii)$ $\alpha=\prod_{m=1}^M{I_m^t}$, $(iii)$ $X_n^{t+1}\in\mathcal{W}(\beta)$, $(iv)$ $X_n^{t+1}R_n^t=R_n^tX_n^t$, $(v)$ $I_{n,m}^t>0$.
\end{prop}

\proof Let us recall the inequality: $\alpha>\beta>0$. Denote $X_\alpha:=X_n^t\vert_{y=(-1)^M\alpha}$, $S_\alpha:=S\vert_{y=(-1)^M\alpha}$ and $V_n=\mathrm{diag}(V_{n,1}^t,\dots,V_{n,M}^t)$. We start with (\ref{eq:a.8}): $\kappa^{-1}\vect{\mu}=X_\alpha^{-1}\vect{\mu}$. It is sufficient to prove the existence and the uniqueness of a real eigenvector $\vect{\mu}=(\mu_1,\dots,\mu_M)^T$ such that $(-1)^{k+1}\mu_k>0$ for all $k$. (Note that $I_{n,i}^t=-\mu_{i+1}/\mu_i$). The vector $\vect{\mu}$ must satisfy $P\vect{\mu}>0$, where $P=\mathrm{diag}(1,-1,1,\dots,(-1)^M)$. By (\ref{eq:11}), we have
\begin{align*}
X_\alpha^{-1}=(E+S_\alpha^{-1}V_n)^{-1}S_\alpha^{-1}
(E+S_\alpha^{-1}V_{n+1})^{-1}S_\alpha^{-1}\cdots
(E+S_\alpha^{-1}V_{n+N-1})^{-1}S_\alpha^{-1}.
\end{align*}
By the inequality $\alpha>\beta>0$, $V_{n,m}^t>0$, we have the following expression:
\[
(E+S^{-1}_\alpha V_{n'})^{-1}=E-S^{-1}_\alpha V_{n'}+
(S^{-1}_\alpha V_{n'})^2-(S^{-1}_\alpha V_{n'})^3
+(S^{-1}_\alpha V_{n'})^4-\cdots,
\]
which implies that the $(i,j)$-element of $(E+S^{-1}_\alpha V_{n'})^{-1}$ is positive (\textit{resp.} negative) if $i+j$ is even (\textit{resp.}\,odd). In other words, $P(E+S^{-1}_\alpha V_{n'})^{-1} P$ is a positive matrix. Therefore $(-1)^NPX_\alpha^{-1}P$ is also a positive matrix. The proposition follows from this fact and the Perron-Frobenius theorem. $\qed$

\subsection{Proof of lemma \ref{lem:2.5}}

In this section, we calculate the determinants of matrices introduced in \S \ref{sec:2.3}. First, the equations $\det{R_n^t}=\alpha-(-1)^My$, $\det{L_n^t}=\beta-(-1)^My$ are straightforward.

Next, we recall the equation (\ref{eq:21}). The $z_{i,j}$ in (\ref{eq:21}) should satisfy
\[
z_{i,j}+yz_{i,j+M}+y^2z_{i,j+2M}+\cdots=\{(i,j)\mbox{-th element of } 
X_{n,m}^t(y)\}.
\]
Define two vectors $\vect{I}$ and $\vect{\mu}$ by 
\begin{gather*}
\textstyle\vect{I}=(1,-I_{n,1}^t,I_{n,1}^tI_{n,2}^t,\dots,\prod_{m=1}^{N-1}(-I_{n,m}^t))^T,\quad
\textstyle\vect{\mu}=(1,-I_{n,1}^t,I_{n,1}^tI_{n,2}^t,\dots,\prod_{m=1}^{M-1}(-I_{n,m}^t))^T. 
\end{gather*}
Then, we have
\[
(z_{1,1},z_{1,2},\dots,z_{1,N})\cdot \vect{I}=
(1,0,0,\dots,0)\cdot (X_{n,m}^t\vert_{y=(-1)^M\alpha})\cdot\vect{\mu}.
\]
By (\ref{eq:a.8}), this equals to $\kappa$. We hence have  $M_{n,m}^{t}\vect{I}=(0,\dots,0,x-\kappa)^T$. On the other hand, due to the definition of $M_n^t$, $\det{M_n^t}$ must be a monic or anti-monic polynomial in $x$ of degree $1$. Therefore, we have $\det{M_n^t}=(-1)^{N+1}(x-\kappa)$.

We calculate $\det{H_n^t}$ similarly. Let 
\begin{gather*}
\textstyle\vect{V}=(1,-V_{n,1}^t,V_{n,1}^tV_{n,2}^t,\dots,\prod_{m=1}^{N-1}(-V_{n,m}^t))^T, \quad
\textstyle\vect{\nu}=(1,-V_{n,1}^t,V_{n,1}^tV_{n,2}^t,\dots,\prod_{m=1}^{M-1}(-V_{n,m}^t))^T. 
\end{gather*}
Then, we have
\[
(z_{1,1},z_{1,2},\dots,z_{1,N})\cdot \vect{V}=
(1,0,0,\dots,0)\cdot (X_{n,m}^t\vert_{y=(-1)^M\beta})\cdot\vect{\nu}=0,
\]
which implies $H_{n,m}^t\vect{V}=(0,\dots,0,x)^T$. Therefore, $\det{H_{n,m}^t}=(-1)^{N+1}x$.

At last, we can directly calculate
$\det{U_m^t}=(-1)^M(x-z_{1,1})=(-1)^{N+1}(x-V_{N,m}^{t}\cdots V_{2,m}^{t}V_{1,m}^{t})$.
$\qed$

\end{document}